\PassOptionsToPackage{pdfpagelabels=false}{hyperref}
\documentclass[useAMS,usenatbib]{mnras}

\def\spose#1{\hbox to 0pt{#1\hss}}
\def\ltsim{$\mathrel{\spose{\lower 3pt\hbox{$\sim$}}
        \raise 2.0pt\hbox{$<$}}$\thinspace}
\def\gtsim{$\mathrel{\spose{\lower 3pt\hbox{$\sim$}}
        \raise 2.0pt\hbox{$>$}}$\thinspace}

\def \deg {$^\circ$}
\def \eg {e.g.}

\def \ie {i.e.}

\newcommand{\chandra }{{\em Chandra}}

\usepackage{graphicx}
\usepackage{amssymb}

\title[A shock at the radio relic position in Abell 115]{A shock at the radio 
relic position in Abell 115}

\author[Botteon et al.]{A.~Botteon$^{1,2}$\thanks{E-mail: 
botteon@ira.inaf.it}, 
F.~Gastaldello$^{3}$, G.~Brunetti$^{2}$ and D.~Dallacasa$^{1,2}$ \\
$^{1}$Dipartimento di Fisica e Astronomia, Universit\`{a} di Bologna, via 
C.~Ranzani 1, I-40127 Bologna, Italy \\
$^{2}$INAF - ORA, via P.~Gobetti 101, I-40129 Bologna, Italy\\
$^{3}$INAF - IASF Milano, via E.~Bassini 15, I-20133 Milano, Italy\\}
\date{\today}

\date{Accepted XXX. Received YYY; in original form ZZZ}

\pubyear{2016}

\begin{document}
\label{firstpage}
\pagerange{\pageref{firstpage}--\pageref{lastpage}}
\maketitle

\begin{abstract}
We analyzed a deep \chandra\ observation ($334\:\rm{ks}$) of the galaxy cluster 
Abell 115 and detected a shock co-spatial with the radio relic. The 
X-ray surface brightness profile across the shock region presents a 
discontinuity, corresponding to a density compression factor 
$\mathcal{C}=2.0\pm0.1$ leading to a Mach number $\mathcal{M}=1.7\pm0.1$ 
($\mathcal{M}=1.4-2$ including systematics). 
Temperatures measured in the upstream and downstream regions are consistent 
with what expected for such a shock: $T_u=4.3^{+1.0}_{-0.6}\:\rm{keV}$ and 
$T_d=7.9^{+1.4}_{-1.1}\:\rm{keV}$ respectively, implying a Mach number 
$\mathcal{M}=1.8^{+0.5}_{-0.4}$. So far, only few other shocks discovered in 
galaxy clusters are consistently detected from both density and temperature 
jumps. The spatial coincidence between this discontinuity and the radio relic 
edge strongly supports the view that shocks play a crucial role in powering 
these synchrotron sources. We suggest that the relic is originated by shock 
re-acceleration of relativistic electrons rather than acceleration from the 
thermal pool. The position and curvature of the shock and the associated relic 
are consistent with an off-axis merger with unequal mass ratio where the shock 
is expected to bend around the core of the less massive cluster. 
\end{abstract}

\begin{keywords}
shock waves -- X-rays: galaxies: clusters -- galaxies: clusters: individual: 
A115 -- radio continuum: general -- radiation mechanisms: non-thermal
\end{keywords}

\section{Introduction}

Galaxy clusters hierarchically form by the aggregation of smaller structures. 
In the process of cluster formation, shocks and turbulence are produced in the 
intra-cluster medium (ICM) and dissipate a significant fraction of kinetic 
energy. A fraction of the dissipated energy can also be channelized into 
non-thermal components, such as magnetic fields and relativistic particles 
\citep[\eg][for a review]{brunetti14rev}. \\
Shock waves are detectable in the X-rays and appear as sharp surface 
brightness discontinuities related to a temperature increase in the 
downstream region. However, shocks in galaxy clusters are difficult to detect 
because projections effects could hide temperature and density jumps and 
because the strongest shocks are expected in the clusters outskirts, where the 
X-ray brightness is low. For this reason, only few of them have been clearly 
detected, \ie\ showing both surface brightness and temperature jumps 
\citep[\eg][]{markevitch02bullet, markevitch05, russell10, macario11, 
dasadia16a665}. \\
There is a broad consensus that shocks play an important role in the 
(re)acceleration of radio emitting electrons generating radio relics 
\citep[see][for reviews]{bruggen12rev, brunetti14rev}. In general, radio relics 
are polarized synchrotron sources found in the periphery of a number of galaxy 
clusters \citep[\eg][for an observational overview]{feretti12rev}. They have 
arc-shaped morphologies and, in some cases, their connection with shocks waves 
is directly established by X-ray observations \citep[\eg][]{macario11, 
akamatsu13systematic, bourdin13}. \\
\indent
Abell 115 (hereafter A115) is an X-ray luminous ($L_X = 1.5 \times 
10^{45}\:\rm{erg\,s^{-1}}$ in the 0.1-2.4$\:\rm{keV}$ band) and dinamically 
disturbed cluster at $z=0.197$. Early X-rays observations 
\citep{forman81, shibata99} revealed that A115 has a X-ray brightness 
distribution characterized by two peaks, the brightest one being in the North 
region and roughly coincident with the position of the radiogalaxy 3C28. More 
recently, \chandra\ observations \citep{gutierrez05} suggested that A115 is 
undergoing an off-axis merger, while optical studies confirmed the presence 
of two sub-clusters in a merging state \citep{barrena07}. In the radio band, 
A115 exhibits a giant radio relic at the edge of the northern part of the 
cluster that extends for $\sim1.5\:\rm{Mpc}$ \citep{govoni01six}. \\
\indent
In this Letter we report the discovery of a shock associated with the radio 
relic from the analysis of \chandra\ and VLA observations of A115. Hereafter, 
we assume a concordance $\Lambda$CDM cosmology with $H_0 = 
70\:\rm{km\,s^{-1}\,Mpc^{-1}}$, $\Omega_{m}=0.3$ and $\Omega_{\Lambda}=0.7$. 
At the redshift of A115 ($z=0.197$), the luminosity distance is 
$673\:\rm{Mpc}$ and $1''$ corresponds to $3.261\:\rm{kpc}$. 
Reported uncertainties are 68\%, unless stated otherwise.

\section{Observations and data reduction}

\subsection{X-ray data reduction}

\begin{figure}
 \centering
 \includegraphics[width=\hsize]{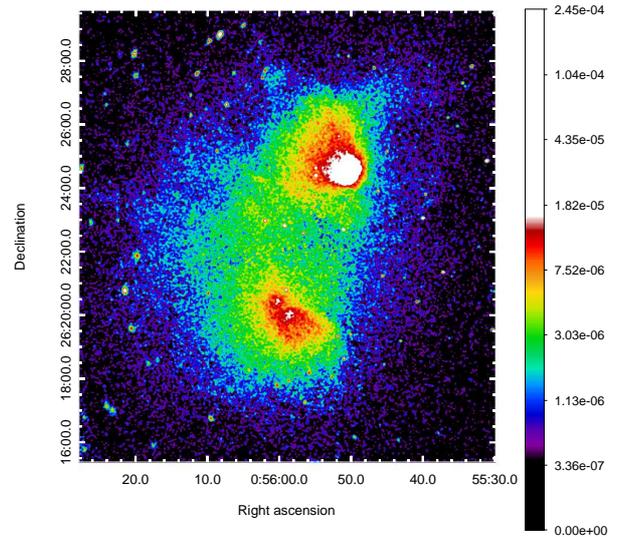}
 \caption{\chandra\ exposure-corrected image in the 0.5-2$\:\rm{keV}$ band of 
A115. The image is smoothed on a $3''$ scale. The colorbar is in 
logarithmic scale, units are $\rm{count\,s^{-1}}$.}
 \label{fig:cluster_only}
\end{figure}

We analyzed the \chandra\ ACIS-I observations of A115 in \texttt{VFAINT} mode 
(ObsID: 3233, 13458, 13459, 15578, 15581) with \texttt{CIAO 4.7} and 
\chandra\ \texttt{CALDB 4.6.9}. All data were reprocessed from 
\texttt{level=1} event file following the standard \chandra\ reduction threads. 
For the observations in which the S3 chip was active (ObsID: 3233, 13459, 
15581), we extracted light curves from this chip in the 0.5-2$\:\rm{keV}$ band 
and we cleaned from soft proton flares using the \texttt{deflare} task with the 
\texttt{clean=yes} option. For the other ObsIDs, light curves were instead 
extracted from a cluster free emission region in one ACIS-I chip. We combined 
the observations with the \texttt{merge\_obs} script and produced the 
0.5-2$\:\rm{keV}$ image binned by a factor of 2 shown in 
Fig.~\ref{fig:cluster_only}. The total cleaned exposure time of this image is 
$334\:\rm{ks}$. \\
For each observation we created a point spread function (PSF) map at 
$1.5\:\rm{keV}$. These were combined with the corresponding exposure 
maps in order to obtain a single exposure-corrected PSF map with minimum 
size. We then ran the \texttt{wavdetect} task on the merged image in order to 
detect point sources. Sources were visually confirmed and then excluded in the 
further analysis. We used the task \texttt{reproject\_event} to match 
background templates to the corresponding event files for every ObsID. Then they 
were normalized by counts in the 9.5-12$\:\rm{keV}$ energy band and combined in 
a single background image subtracted in the surface brightness analysis. \\
\indent
Spectral extraction was performed for every ObsID using the same regions. We 
modeled the particle background following \cite{bartalucci14} and the sky 
component with two thermal plasmas at temperatures of $0.14$ and 
$0.25\:\rm{keV}$ to account for the Galactic emission and an absorbed 
power-law with photon index $\Gamma=1.4$ to account for the cosmic X-ray 
background. The background parameters were determined by fitting 
spectra extracted from cluster-free emission regions at the edge of the field of 
view in the 0.5-11$\:\rm{keV}$ energy band. The cluster emission was fitted 
with a thermal model with metal abundance fixed at $0.3\:\rm{Z_\odot}$. All 
fits were performed using Cash statistics. The robustness of the fits was 
verified by checking for systematic errors due to the background determination. 
For this reason we re-performed the cluster spectral analysis with the 
background normalization levels fixed at $\pm1\sigma$ within their best fit 
values. In addition, the impact of ACIS QE contamination at low energy was 
verified by fitting in the bands 0.5-11 and 1-11$\:\rm{keV}$. In both cases, 
the thermal parameters of the fits are consistent within $1\sigma$. 

\subsection{Radio data reduction}

\begin{figure*}
 \centering
 \includegraphics[width=\textwidth]{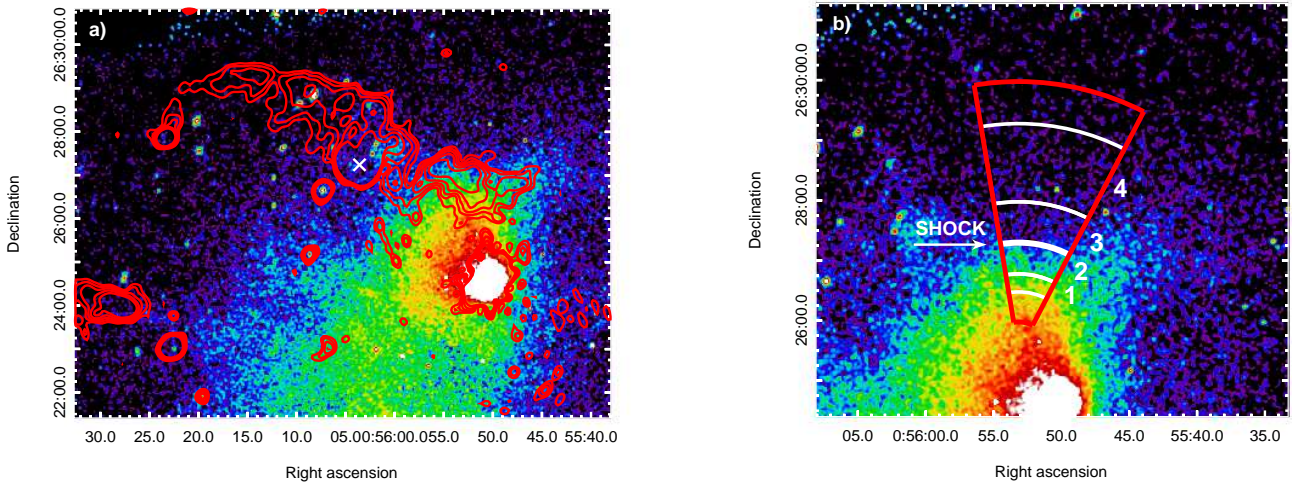}
 \caption{Zoom on the northern region of A115. \textit{a}) Radio contours at 
$1.4\:\rm{GHz}$ of the relic source in A115. The resolution is $15'' 
\times 14''$. Contour levels are $3\sigma \times 
\pm(\sqrt{2})^n\:\rm{mJy\,b^{-1}}$, with $n=1,2,3,4$. The $1\sigma$ noise 
level is $70\:\rm{\mu Jy\,b^{-1}}$. White cross denotes the $0056+26\:\rm{B}$ 
radiogalaxy. \textit{b}) Red sector delineates the region where the X-ray 
surface brightness profile was extracted; white sectors represent the four bins 
where spectral analysis was performed. In both panels colors represent the 
\chandra\ image (same of Fig.~\ref{fig:cluster_only}).}
 \label{fig:all_cluster}
\end{figure*}

We re-analyzed VLA archival data at $1.4\:\rm{GHz}$ in the C and D 
configurations. Details of observations can be found in \cite{govoni01six}. The 
two dataset were edited, calibrated and imaged separately. Particular care was 
devoted to the identification and removal of bad data corrupted by intermittent 
radio frequency interference. After a number of phase-only 
self-calibration iterations and an accurate comparison of the flux densities of 
the sources, the data of IF1, at $1364\:\rm{MHz}$ for both datasets, were 
combined (the second IF could not be used given the significantly different 
frequency). \\
The final dataset ($4.5\:\rm{h}$ and $\Delta\nu=50\:\rm{MHz}$) was once again 
self-calibrated a number of times and final images were obtained using 
different weighting schemes. The image shown in Fig.~\ref{fig:all_cluster}a was 
made with a two-scale clean where the extended emission was deconvolved 
using a larger beam ($\sim 30''$). The restoring beam is $15'' \times 14''$ 
in position angle $-35$\deg\ and the off-source noise level is 
$70\:\rm{\mu Jy\,b^{-1}}$. Errors on flux densities are dominated by the 5\% 
uncertainty of the absolute flux scale calibration. 

\section{Results}

\begin{figure}
 \centering
 \includegraphics[width=\hsize]{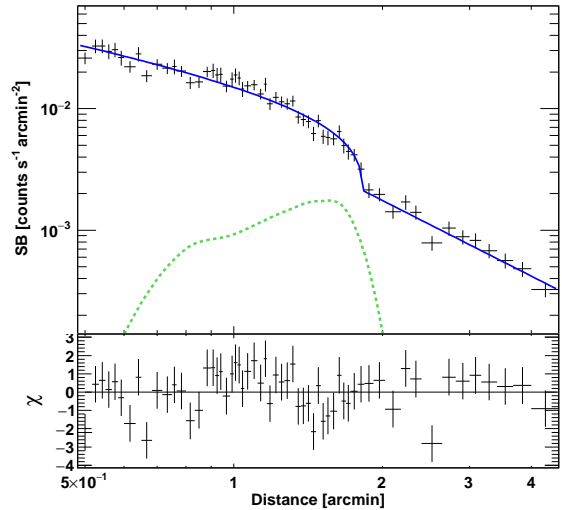}
 \caption{X-ray surface brightness profile in the 0.5-2$\:\rm{keV}$ band 
extracted in the red region shown in Fig.~\ref{fig:all_cluster}b. The data were 
rebinned to reach a minimum signal-to-noise ratio of 7. The fit had 
$\chi^2 / \rm{d.o.f.} = 1.2$. The green dashed line shows the radio 
relic brightness profile (in arbitrary units).}
 \label{fig:sb_profile}
\end{figure}

\begin{figure}
 \centering
 \includegraphics[width=.8\hsize]{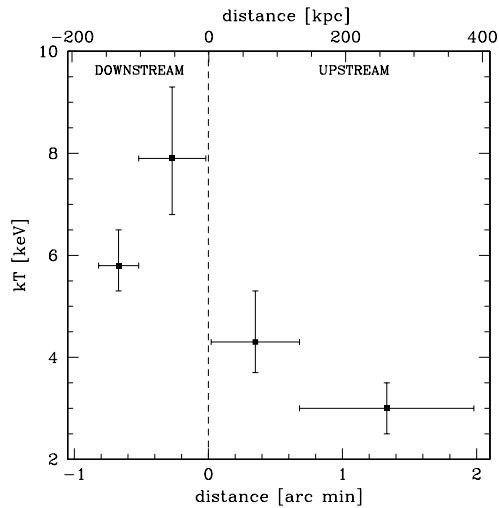}
 \caption{Temperature profile across the shock. The vertical dashed line sets 
the location of the X-ray surface brightness discontinuity.}
 \label{fig:T_profile}
\end{figure}

A visual inspection of the X-ray image of A115 led us to identify a surface 
brightness jump in the northern part of the system, co-spatially located with 
the relic position (Fig.~\ref{fig:all_cluster}). The surface brightness profile 
was extracted in the red sector shown in Fig.~\ref{fig:all_cluster}b and 
fitted with \texttt{PROFFIT v1.3} \citep{eckert11}. The data were fitted 
assuming a underlying broken power-law density profile in the form

\begin{equation}\label{eq:break-pl}
 \begin{array}{ll}
 n_d (r) = \mathcal{C} n_0 \left( \frac{r}{r_{sh}} \right)^{a_1}, & 
\mbox{if} \quad r \leq r_{sh} \\
\\
 n_u (r) = n_0 \left( \frac{r}{r_{sh}} \right)^{a_2}, & \mbox{if} \quad r 
> r_{sh}
 \end{array}
\end{equation}

\noindent
where $\mathcal{C} \equiv n_d/n_u$ is the compression factor of the 
shock, $n_0$ is the density normalization, $a_1$ and $a_2$ are the power-law 
indices, $r$ is the radius from the center of the sector and $r_{sh}$ is the 
radius corresponding to the putative shock front (which curvature relative to 
the line-of-sight is assumed to be the same as that in the plane of the sky). 
Subscripts $u$ and $d$ indicate quantities upstream and downstream of the 
shock, respectively. All the parameters of the model were let free in 
the fit. The best fit broken power-law is shown in Fig.~\ref{fig:sb_profile}. 
The compression factor taken from the red sector in Fig.~\ref{fig:all_cluster}b 
is $\mathcal{C}=2.0\pm0.1$. By using an adiabatic index $\gamma=5/3$ and the 
Rankine-Hugoniot jump conditions this leads to a Mach number 
$\mathcal{M}=1.7\pm0.1$. Obviously this value does not include 
systematics deriving from 3D model geometry and the shape of the extraction 
region. We explored uncertainties due to the sector choice by varying its 
curvature radius, aperture and position angle. Tests were made keeping the 
discontinuity distance frozen. Changing the shock curvature radius from its 
best value $360\:\rm{kpc}$ by a factor 0.5 and 1.5 gives the highest variation 
in terms of the compression factor, $1.6-2.1$, corresponding to 
$\mathcal{M}=1.4-1.8$. Varying the other parameters of the 
region results in values of $\mathcal{M}$ within this range. We did not 
introduce any ellipticity in the problem  as the surface brightness edge looks 
quite straight. The red sector in Fig.~\ref{fig:all_cluster}b represents the 
best compromise to highlight the discontinuity in terms of $\chi^2 / 
\rm{d.o.f.}$. \\
\indent
In a shock wave, the downstream region is characterized by a temperature 
increase; in a cold front instead the denser region has a lower temperature 
\citep[see \eg][]{markevitch07rev}. For this reason, we extracted spectra in 
the four white regions shown in Fig.~\ref{fig:all_cluster}b and performed 
spectral fitting. We found a temperature jump from 
$T_u=4.3^{+1.0}_{-0.6}\:\rm{keV}$ to $T_d=7.9^{+1.4}_{-1.1}\:\rm{keV}$ in 
the two bins closer to the surface brightness discontinuity, confirming the 
shock nature of this feature. This temperature jump corresponds to 
$\mathcal{M}=1.8^{+0.5}_{-0.4}$, in agreement with the Mach number derived from 
the density jump; projection effects, if they play a role, are expected to make 
the intrinsic jump slightly smaller. The temperature profile taken across the 
shock region is shown in Fig.~\ref{fig:T_profile}. The first bin 
exhibits a low temperature, this could either be the result of the gas expansion 
behind the shock or the presence of a substructure along the line-of-sight 
\citep[\eg][]{markevitch02bullet}. For a sanity check we re-performed the 
surface brightness fit avoiding this temperature bin by excluding data points 
at $r<1.5'$; in this case we achieve $\mathcal{C}=2.0 \pm 0.3$, leading to 
$\mathcal{M}=1.7^{+0.3}_{-0.2}$.

\section{Discussion}

\subsection{Radio relic--shock connection}\label{ch:connection}

At the resolution of $15'' \times 14''$, the radio relic in A115 presents a 
discontinuity in the center of its northern edge structure, roughly splitting 
it in two parts (Fig.~\ref{fig:all_cluster}a). One is quite short, straight in 
the E-W direction and coincides with the shock location, while the remainder 
extends in the eastern direction beyond the cluster X-ray emission and appears 
slightly bent. Given the spatial coincidence with the shock we suggest that the 
former is a ``classical'' radio relic where particles are accelerated or 
re-accelerated by the passage of the shock. In this restricted region, the 
radio flux density at $1.4\:\rm{GHz}$ is $F = 34 \pm 2\:\rm{mJy}$. The eastern 
radio emission is more difficult to interpret and it could be produced by the 
uplifting of the plasma coming from the cluster radio bright source 
$0056+26\:\rm{B}$ (a narrow angle tailed radiogalaxy whose emission fades away 
in the East direction), located in the middle of the relic, after the shock 
passage. \\
\indent
In a case of a head-on merging in the plane of the sky between two clusters 
with similar mass, radio relics are expected to come in pairs in opposite 
directions along the axis merger \citep[\eg][]{rottgering97}. This is clearly 
not the case of A115 where the relic has an unusual location as it is 
oriented almost parallel to the northern sub-cluster motion. This led to doubts 
about the nature of relic extended emission and to the interpretation as tails 
of radio plasma trailing the radiogalaxies \citep{gutierrez05}. However, 
numerical simulations of an off-axis merger between clusters with different 
mass predicts that the shock curves around the core of the minor cluster 
\citep[see Fig.~7 in][]{ricker01}, in agreement with A115-N sub-cluster being 
less massive \citep{barrena07}. 

\subsection{Acceleration efficiency}

\begin{figure}
 \centering 
 \includegraphics[width=.8\hsize]{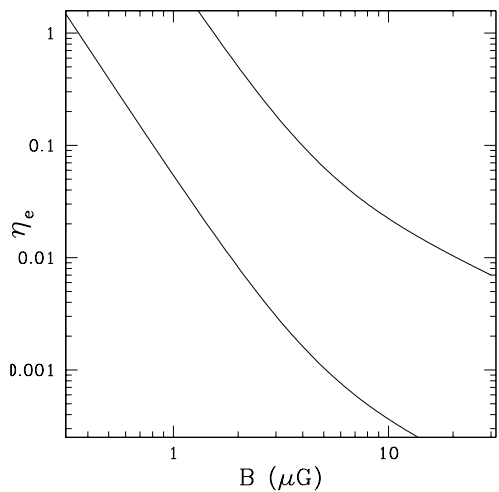}
 \caption{Electron acceleration efficiency versus magnetic field downstream in 
the shock in A115. Lines represent efficiencies evaluated for 
re-accelerated electrons with $p_{min}/m_e c=20$ (top) and 200 (bottom) and 
$\delta_{inj}=3.8$.}
 \label{fig:eff}
\end{figure}

There is consensus on the hypothesis that shocks play an important role for the 
origin of radio relics. However the case of shock acceleration of cosmic ray 
electrons (CRe) from the thermal ICM is challenged by the large efficiencies 
that are required to reproduce the total radio luminosity of several relics 
\citep[][for review]{brunetti14rev}. Large acceleration efficiencies would also 
lead to a cosmic ray protons (CRp) induced $\gamma$-ray emission exceeding 
current clusters upper limits in the case that shocks in the ICM produce a 
ratio CRp/CRe similar to that in supernova remnants (SNR) 
\citep{vazza14challenge}. To alleviate the large requirements for acceleration 
efficiencies of CRe, recent models for cluster relics assume shock 
re-acceleration of a pre-existing population of CRe \citep{markevitch05, kang12, 
pinzke13, kang14}. \\
\indent
A115 is a test case to constrain the origin of the shock--relic connection, 
because the underlying shock is well constrained. If the downstream synchrotron 
luminosity emitted at frequency $\geq \nu_0$ originates from electrons in 
steady state conditions, the bolometric ($\geq \nu_0$) synchrotron luminosity 
that is generated via shock acceleration from a shock with speed $V_{sh}$, 
surface $S$ and upstream mass-density $\rho_u$ can be estimated as

\begin{equation}\label{eq:luminosity}
\int_{\nu_0} L(\nu)\,d\nu \simeq {1 \over 2 } \eta_e \Psi \rho_u 
V_{sh}^3 \left( 1 - {1 \over{\mathcal{C}^2}} \right) {{ B^2 }\over{B_{cmb}^2 + 
B^2}} S
\end{equation}

\noindent
where $\eta_e$ is the efficiency of electron acceleration,   

\begin{equation}\label{eq:psi}
\Psi = {{\int_{p_{min}} Q(p) E\,dp}\over{\int_{p_0} Q(p) E\,dp}}
\end{equation}

\noindent
accounts for the ratio of the energy flux  injected in ``all'' electrons and 
those visible in the radio band ($\nu \geq \nu_0$), $p_0$ is the momentum of 
the relativistic electrons emitting the synchrotron frequency $\nu_0$ in a 
magnetic field $B$ and $B_{cmb}=3.25(1+z)^2\:\rm{\mu G}$ accounts for inverse 
Compton scattering of cosmic microwave background photons. \\
\indent
A model of shock acceleration from the thermal pool for the relic is 
readily rule out by our measurements. According to diffusive shock acceleration 
theory the particles injection spectrum is $Q(p) \propto p^{-\delta_{inj}}$ 
with $\delta_{inj} = 2 (\mathcal{M}^2+1)/(\mathcal{M}^2-1)$ 
\citep[\eg][]{blandford87rev}, that for $\mathcal{M}=1.7-1.8$ would imply 
$\delta_{inj} = 4-3.8$ (and integrated spectral index $\alpha = \delta_{inj} /2 
= 2-1.9$). Not only this is inconsistent with the measured spectrum of the 
relic ($\alpha \sim 1.1$, \citealt{govoni01six}), but 
from Eq.~\ref{eq:luminosity} it also requires an untenable large acceleration 
efficiency (assuming $S = \pi \times 180 \times 180\:\rm{kpc^2}$).\\ 
Alternatively, we can assume re-acceleration. In this case the initial 
(upstream) and accelerated spectra of electrons are connected via

\begin{equation}\label{eq:n_p}
N_d(p) = (\delta_{inj} +2 ) p^{-\delta_{inj}} \int_{p_{min}}^{p} 
x^{\delta_{inj}-1} N_u(x)\,dx
\end{equation}

\noindent
where $N_u$ refers to the upstream spectrum of seed particles; re-acceleration 
efficiencies can be larger compared to the case of acceleration from thermal 
ICM, because seed ultra-relativistic electrons diffuse efficiently across the 
shock discontinuity \citep[\eg][]{brunetti14rev}. In Fig.~\ref{fig:eff} we show 
the case of a spectrum of re-accelerated electrons with power-law $\delta_{inj} 
= 3.8$ and $p_{min}/m_e c=20$ and 200. Re-acceleration of electrons with 
$p_{min} \geq 100 m_e c$ appears energetically viable. In this case, however, 
the spectrum of the relic would be very steep ($\alpha \sim 2$). Alternatively, 
the shock may re-accelerate a cloud of electrons that are not very old and have 
a flatter spectrum. In this case the shock essentially boosts their emission at 
higher frequencies preserving the seed spectrum (Eq.~\ref{eq:n_p}, see \eg\ 
\citealt{kang16arx}, for details). As already discussed in 
Section~\ref{ch:connection}, re-acceleration is also 
suggested by the morphology of the radio relic and by the fact that the relic 
embeds a few radiogalaxies that would be natural sources of seed particles. 
Finally, we note that the eastern part of the radio relic deploys in a region 
of low X-ray surface brightness where the thermal energy density is small and 
where a scenario of shock acceleration of thermal electrons would require an 
efficiency that is even larger than that shown in Fig.~\ref{fig:eff}.

\section{Conclusions}

We presented results concerning the merging galaxy cluster A115 at 
$z=0.197$. We analyzed a \chandra\ dataset for a total exposure 
time of $334\:\rm{ks}$ and archival VLA radio observations at $1.4\:\rm{GHz}$. 
In this Letter we focused on A115-N, where a giant radio relic stands out. \\
\indent
The deep \chandra\ observations led us to detect a shock spatially coincident 
with the radio relic in A115. Assuming a broken power-law density profile, the 
surface brightness discontinuity is consistent with a density compression factor 
$\mathcal{C}=2.0\pm0.1$, which in turn results in $\mathcal{M}=1.7\pm0.1$; the 
Mach number is in the range $1.4-2$ including systematic uncertainties. The 
shock nature of the discontinuity was confirmed by spectral analysis of the 
downstream and the upstream regions, where a temperature jump from 
$T_d=7.9^{+1.4}_{-1.1}\:\rm{keV}$ to $T_u=4.3^{+1.0}_{-0.6}\:\rm{keV}$ was 
found, implying $\mathcal{M}=1.8^{+0.5}_{-0.4}$. This is one of the few cases 
where the surface brightness and temperature drops in a merger shock are 
clearly detected and are in excellent agreement. \\
\indent
In the radio band, the relic can be roughly divided into a W and an E part. The 
former is a ``classical'' relic spatially coincident with the shock found in 
the X-rays. The relic location is in agreement with an off-axis merger between 
two clusters with unequal mass, where the shock bends around the core of the 
less massive system. The eastern relic radio emission is harder to interpret 
as, in this region, a proper X-ray analysis cannot be performed given its low 
surface brightness. An attractive scenario is the uplifting of the cluster 
radiogalaxy $0056+26\:\rm{B}$ plasma after the shock sweeping in the cluster 
outskirts. \\
\indent
Confirming the presence of a shock underlying a radio relic strongly supports 
the scenario where radio relics are powered by emitting particles produced 
during merger shocks. Given the low Mach number, spectrum and morphology of the 
relic, models in which relativistic seed electrons are re-accelerated by the 
shock passage are favored. In this respect, the few cluster radiogalaxies 
embedded in the relic in A115 could naturally provide the required seed 
particles.  

\section*{Acknowledgments}

We thank the anonymous referee for the useful comments on the manuscript. We 
are grateful to Dominique Eckert for developing and making available a powerful 
tool such as \texttt{PROFFIT}. The scientific results reported in this article 
are based on observations made by the Chandra X-ray Observatory. The NRAO is a 
facility of the National Science Foundation operated under cooperative 
agreement by Associated Universities, Inc. AB and GB acknowledge partial 
support from PRIN-INAF 2014. FG acknowledges financial support from PRIN-INAF 
2012 and ASI-INAF I/037/12/0.

\bibliographystyle{mnras}
\bibliography{biblio}

\begin{thebibliography}{}
\makeatletter
\relax
\def\mn@urlcharsother{\let\do\@makeother \do\$\do\&\do\#\do\^\do\_\do\%\do\~}
\def\mn@doi{\begingroup\mn@urlcharsother \@ifnextchar [ {\mn@doi@}
  {\mn@doi@[]}}
\def\mn@doi@[#1]#2{\def\@tempa{#1}\ifx\@tempa\@empty \href
  {http://dx.doi.org/#2} {doi:#2}\else \href {http://dx.doi.org/#2} {#1}\fi
  \endgroup}
\def\mn@eprint#1#2{\mn@eprint@#1:#2::\@nil}
\def\mn@eprint@arXiv#1{\href {http://arxiv.org/abs/#1} {{\tt arXiv:#1}}}
\def\mn@eprint@dblp#1{\href {http://dblp.uni-trier.de/rec/bibtex/#1.xml}
  {dblp:#1}}
\def\mn@eprint@#1:#2:#3:#4\@nil{\def\@tempa {#1}\def\@tempb {#2}\def\@tempc
  {#3}\ifx \@tempc \@empty \let \@tempc \@tempb \let \@tempb \@tempa \fi \ifx
  \@tempb \@empty \def\@tempb {arXiv}\fi \@ifundefined
  {mn@eprint@\@tempb}{\@tempb:\@tempc}{\expandafter \expandafter \csname
  mn@eprint@\@tempb\endcsname \expandafter{\@tempc}}}

\bibitem[\protect\citeauthoryear{Akamatsu \& Kawahara}{Akamatsu \&
  Kawahara}{2013}]{akamatsu13systematic}
Akamatsu H.,  Kawahara H.,  2013, \mn@doi [PASJ] {10.1093/pasj/65.1.16}, 65, 16

\bibitem[\protect\citeauthoryear{Barrena, Boschin, Girardi  \& Spolaor}{Barrena
  et~al.}{2007}]{barrena07}
Barrena R.,  Boschin W.,  Girardi M.,   Spolaor M.,  2007, \mn@doi [A{\&}A]
  {10.1051/0004-6361:20077407}, 469, 861

\bibitem[\protect\citeauthoryear{Bartalucci, Mazzotta, Bourdin  \&
  Vikhlinin}{Bartalucci et~al.}{2014}]{bartalucci14}
Bartalucci I.,  Mazzotta P.,  Bourdin H.,   Vikhlinin A.,  2014, \mn@doi
  [A{\&}A] {10.1051/0004-6361/201423443}, 566, A25

\bibitem[\protect\citeauthoryear{Blandford \& Eichler}{Blandford \&
  Eichler}{1987}]{blandford87rev}
Blandford R.,  Eichler D.,  1987, \mn@doi [Phys. Rep.]
  {10.1016/0370-1573(87)90134-7}, 154, 1

\bibitem[\protect\citeauthoryear{Bourdin, Mazzotta, Markevitch, Giacintucci  \&
  Brunetti}{Bourdin et~al.}{2013}]{bourdin13}
Bourdin H.,  Mazzotta P.,  Markevitch M.,  Giacintucci S.,   Brunetti G.,
  2013, \mn@doi [ApJ] {10.1088/0004-637X/764/1/82}, 764, 82

\bibitem[\protect\citeauthoryear{Br{\"{u}}ggen, Bykov, Ryu  \&
  R{\"{o}}ttgering}{Br{\"{u}}ggen et~al.}{2012}]{bruggen12rev}
Br{\"{u}}ggen M.,  Bykov A.,  Ryu D.,   R{\"{o}}ttgering H.,  2012, \mn@doi
  [Space Sci. Rev.] {10.1007/s11214-011-9785-9}, 166, 187

\bibitem[\protect\citeauthoryear{Brunetti \& Jones}{Brunetti \&
  Jones}{2014}]{brunetti14rev}
Brunetti G.,  Jones T.,  2014, \mn@doi [Int. J. Mod. Phys. D]
  {10.1142/S0218271814300079}, 23, 30007

\bibitem[\protect\citeauthoryear{Dasadia et~al.,}{Dasadia
  et~al.}{2016}]{dasadia16a665}
Dasadia S.,  et~al., 2016, \mn@doi [ApJ] {10.3847/2041-8205/820/1/L20}, 820,
  L20

\bibitem[\protect\citeauthoryear{Eckert, Molendi  \& Paltani}{Eckert
  et~al.}{2011}]{eckert11}
Eckert D.,  Molendi S.,   Paltani S.,  2011, \mn@doi [A{\&}A]
  {10.1051/0004-6361/201015856}, 526, A79

\bibitem[\protect\citeauthoryear{Feretti, Giovannini, Govoni  \&
  Murgia}{Feretti et~al.}{2012}]{feretti12rev}
Feretti L.,  Giovannini G.,  Govoni F.,   Murgia M.,  2012, \mn@doi [A{\&}A
  Rev.] {10.1007/s00159-012-0054-z}, 20, 54

\bibitem[\protect\citeauthoryear{Forman, Bechtold, Blair, Giacconi, van
  Speybroeck  \& Jones}{Forman et~al.}{1981}]{forman81}
Forman W.,  Bechtold J.,  Blair W.,  Giacconi R.,  van Speybroeck L.,   Jones
  C.,  1981, \mn@doi [ApJ] {10.1086/183459}, 243, L133

\bibitem[\protect\citeauthoryear{Govoni, Feretti, Giovannini, B{\"{o}}hringer,
  Reiprich  \& Murgia}{Govoni et~al.}{2001}]{govoni01six}
Govoni F.,  Feretti L.,  Giovannini G.,  B{\"{o}}hringer H.,  Reiprich T.~H.,
  Murgia M.,  2001, \mn@doi [A{\&}A] {10.1051/0004-6361:20011016}, 376, 803

\bibitem[\protect\citeauthoryear{Gutierrez \& Krawczynski}{Gutierrez \&
  Krawczynski}{2005}]{gutierrez05}
Gutierrez K.,  Krawczynski H.,  2005, \mn@doi [ApJ] {10.1086/426420}, 619, 161

\bibitem[\protect\citeauthoryear{Kang \& Ryu}{Kang \& Ryu}{2016}]{kang16arx}
Kang H.,  Ryu D.,  2016, preprint (\mn@eprint {arXiv} {1602.03278})

\bibitem[\protect\citeauthoryear{Kang, Ryu  \& Jones}{Kang
  et~al.}{2012}]{kang12}
Kang H.,  Ryu D.,   Jones T.,  2012, \mn@doi [ApJ]
  {10.1088/0004-637X/756/1/97}, 756, 97

\bibitem[\protect\citeauthoryear{Kang, Petrosian, Ryu  \& Jones}{Kang
  et~al.}{2014}]{kang14}
Kang H.,  Petrosian V.,  Ryu D.,   Jones T.,  2014, \mn@doi [ApJ]
  {10.1088/0004-637X/788/2/142}, 788, 142

\bibitem[\protect\citeauthoryear{Macario, Markevitch, Giacintucci, Brunetti,
  Venturi  \& Murray}{Macario et~al.}{2011}]{macario11}
Macario G.,  Markevitch M.,  Giacintucci S.,  Brunetti G.,  Venturi T.,
  Murray S.,  2011, \mn@doi [ApJ] {10.1088/0004-637X/728/2/82}, 728, 82

\bibitem[\protect\citeauthoryear{Markevitch \& Vikhlinin}{Markevitch \&
  Vikhlinin}{2007}]{markevitch07rev}
Markevitch M.,  Vikhlinin A.,  2007, \mn@doi [Phys. Rep.]
  {10.1016/j.physrep.2007.01.001}, 443, 1

\bibitem[\protect\citeauthoryear{Markevitch, Gonzalez, David, Vikhlinin,
  Murray, Forman, Jones  \& Tucker}{Markevitch
  et~al.}{2002}]{markevitch02bullet}
Markevitch M.,  Gonzalez A.,  David L.,  Vikhlinin A.,  Murray S.,  Forman W.,
  Jones C.,   Tucker W.,  2002, \mn@doi [ApJ] {10.1086/339619}, 567, L27

\bibitem[\protect\citeauthoryear{Markevitch, Govoni, Brunetti  \&
  Jerius}{Markevitch et~al.}{2005}]{markevitch05}
Markevitch M.,  Govoni F.,  Brunetti G.,   Jerius D.,  2005, \mn@doi [ApJ]
  {10.1086/430695}, 627, 733

\bibitem[\protect\citeauthoryear{Pinzke, Oh  \& Pfrommer}{Pinzke
  et~al.}{2013}]{pinzke13}
Pinzke A.,  Oh S.,   Pfrommer C.,  2013, \mn@doi [MNRAS]
  {10.1093/mnras/stt1308}, 435, 1061

\bibitem[\protect\citeauthoryear{Ricker \& Sarazin}{Ricker \&
  Sarazin}{2001}]{ricker01}
Ricker P.,  Sarazin C.,  2001, \mn@doi [ApJ] {10.1086/323365}, 561, 621

\bibitem[\protect\citeauthoryear{R{\"{o}}ttgering, Wieringa, Hunstead  \&
  Ekers}{R{\"{o}}ttgering et~al.}{1997}]{rottgering97}
R{\"{o}}ttgering H.,  Wieringa M.,  Hunstead R.,   Ekers R.,  1997, \mn@doi
  [MNRAS] {10.1093/mnras/290.4.577}, 290, 577

\bibitem[\protect\citeauthoryear{Russell, Sanders, Fabian, Baum, Donahue, Edge,
  McNamara  \& O'Dea}{Russell et~al.}{2010}]{russell10}
Russell H.,  Sanders J.,  Fabian A.,  Baum S.,  Donahue M.,  Edge A.,  McNamara
  B.,   O'Dea C.,  2010, \mn@doi [MNRAS] {10.1111/j.1365-2966.2010.16822.x},
  406, 1721

\bibitem[\protect\citeauthoryear{Shibata, Honda, Ishida, Ohashi  \&
  Yamashita}{Shibata et~al.}{1999}]{shibata99}
Shibata R.,  Honda H.,  Ishida M.,  Ohashi T.,   Yamashita K.,  1999, \mn@doi
  [ApJ] {10.1086/307819}, 524, 603

\bibitem[\protect\citeauthoryear{Vazza \& Br{\"{u}}ggen}{Vazza \&
  Br{\"{u}}ggen}{2014}]{vazza14challenge}
Vazza F.,  Br{\"{u}}ggen M.,  2014, \mn@doi [MNRAS] {10.1093/mnras/stt2042},
  437, 2291

\makeatother
\end{thebibliography}

\bsp	
\label{lastpage}
\end{document}